\documentclass[10pt,a4paper,twocolumn]{article}
\usepackage{bachelorproject}
\usepackage{booktabs}
\usepackage{hyperref}
\usepackage{natbib}

\title{\vspace{-0.5in}{\bfseries\scshape 
    Identifying Suspected Mislabeled Apps in Google Play Application Removal Prediction: An Empirical Comparison of Label Noise Detection Methods
    } \\ 
    \vspace{1ex} 
}

\author{
    \normalsize{Deborah Dobles Montalvan, University of Groningen} \\ 
    \normalsize{Co-Authors: F. Mohsen \& H. de Weerd}
}

\date{\vspace{-5ex}}

\begin{document}

\twocolumn[
  \maketitle
  \begin{@twocolumnfalse}
    \begin{abstract}

Models that predict which Google Play apps will be removed are trained on labels that
record only whether an app was still in the store at a later observation. An
app that has disappeared is labeled \emph{removed} and one that is still present is
labeled \emph{stable}, but neither label records \emph{why}. A developer who withdraws an
app voluntarily and a developer whose app is taken down for a policy violation
both produce the \emph{removed} label, and a spam app that has not yet been caught
keeps the \emph{stable} label. This work calls that mismatch, between the recorded market
status and the policy outcome the model is meant to learn, label noise. Three detectors of
label noise, each from a different methodological family, are applied to the 870,514
apps of \citet{mohsen2022}: Isolation Forest, which looks for apps that are
unusual in the feature space, Neighborhood Disagreement, which looks for apps whose
nearest neighbors carry the opposite label, and Prediction Inconsistency, which looks for
apps a classifier labels differently from the data. The apps flagged by all
three at once are called the overlap, 7,598 of them at the detectors' default settings, and
are treated as the strongest mislabeling candidates. Two questions follow: first, does
removing the flagged apps improve the removal prediction model? It does not. No
detector, and neither the overlap nor the union of the three, improves on the baseline, and
the loss grows with the number of apps removed. Second, are the flagged apps
under-represented among apps whose label VirusTotal and Quark Engine independently
confirm? Among confirmed removals they are, increasingly so as the confirmation threshold
tightens, falling to 0.43 times the expected rate, among confirmed stable apps an
apparent excess disappears once the age profile of the scanned apps is accounted for.
A model trained on only the 3,021 trainable partition overlap apps that fall inside the training data
reaches a test AUC of 0.2518, far below chance, which means the relationship between features
and labels inside that set runs opposite to the rest of the data. A comparison against
unflagged apps carrying the same label shows the inconsistency runs in both
directions: abandoned, spam-like apps carry the stable label while healthy-looking
apps carry the removed label. The contribution is therefore not a cleaning method but
a characterization. The detectors locate and describe a small set of label noise candidates
that they cannot profitably remove.
\end{abstract}

  \end{@twocolumnfalse}
]

\thispagestyle{firststyle}

\section{Introduction}

Google Play hosts millions of apps and removes large numbers of them for violating its policies. However, these removals bring consequences, as users can be exposed to harmful apps of low quality before they are taken down, and developers can lose their work and revenue. It is beneficial to predict which of these apps will be removed for a developer to understand what causes an app to pose risk, but also to support moderation of the store.

Consider three apps: the first is a small game whose developer loses interest and takes it down. The second is a wallpaper app that floods its users with advertisements and is taken down by Google for violating the developer policy. Both disappear from the store between the two observations that produced this dataset, and both therefore carry the label \emph{removed}. The third is a spam clone that has not yet been caught, it is still present at the second observation and carries the label \emph{stable}. A model trained on these labels is asked to predict policy enforcement, but what it is shown is market status. The label is an accurate record of what happened to the app in the store but an unreliable record of why it happened. This work asks which apps carry labels that are inconsistent in this way, and whether a removal prediction model improves once they are taken out.

\citet{mohsen2022} approached this as a prediction problem by building an XGBoost model that predicts whether an app will be removed, using only metadata features such as permissions, rating, and description lengths. This model was trained on 870{,}514 apps and reached an Area Under the Curve (AUC) of 0.792 in the setting centered on the user. \citet{mohsen2022} also treat the removed and stable labels as the ground truth.

These labels are produced by observing whether an app remains present in the store over time. An app that disappears is labeled as removed, whereas an app that remains is labeled as stable. This feature records an app's market status but it does not reflect why it has that status. This status may stem from other reasons that are unrelated to a policy violation such as a voluntary withdrawal by its developer. In contrast, according to \citet{wang2018}, an app that violates a policy could still remain in the store for some time. Therefore, an app's market status could function as a noisy proxy for policy enforcement, since the labels do not consistently reflect whether an app was removed for a violation or other reasons. In this work, the label noise is described as those instances where the given label of either ``removed'' or ``stable'' does not align with the policy status of the app. 

Label noise should not go under the radar. According to \citet{frenay2014}, it has been shown that label noise reduces model accuracy, increases model complexity and increases the amount of training data required to reach a given level of performance. Moreover, pervasive label errors have been found across the most used datasets \citep{northcutt2021}. If these datasets carry errors, a model that is being trained on them will be partly learning from labels that are not entirely correct.

Although label noise detection has been applied to image benchmarks \citep{northcutt2021} and to Android malware datasets \citep{malwhiteout2022}, it has not been applied to the prediction of app removal. Within removal prediction itself, these labels have been treated as correct. Neither Mohsen et al. nor related work has examined whether an app's market status consistently represents whether an app was subject to a policy enforcement. Therefore, this work addresses that gap, where label noise detection is applied to data on app removal.

This leads to the research question of this work: \emph{can label noise detectors
identify apps in the Google Play removal dataset of \citet{mohsen2022} whose status
label is inconsistent with policy enforcement, and does removing those apps improve
the removal prediction model?} The question is deliberately scoped. It concerns one dataset,
three detectors (Isolation Forest, Neighborhood Disagreement and Prediction Inconsistency),
and it is not a claim about tabular datasets in general. It is also not a search for the
detector with the highest accuracy, because this dataset provides no verified reason for any
removal, so no detector output can be scored against a known answer. What can be measured is
whether the flagged apps behave as mislabeled apps would, and whether
removing them helps the model. The question is therefore answered in two parts:

\begin{itemize}
  \item \textbf{Validation 1.} Does removing the apps flagged by a detector improve
        the removal prediction model, compared with a baseline trained on the uncleaned data?
  \item \textbf{Validation 2.} Are the flagged apps under-represented among the
        apps whose label VirusTotal and Quark Engine independently confirm?
\end{itemize}

Because there is no ground truth for the removal reasons, the flagged apps are reported as label noise candidates rather than as confirmed errors throughout. To address this, three label noise detectors that were gathered from different methodological families are applied to the dataset of \citet{mohsen2022}. Their agreement is analyzed and the apps they flag are removed, relabeled, or assessed individually on the model. 

This work makes three contributions.
 
\begin{enumerate}
  \item It applies label noise detection to app removal prediction, which has not
        been done before in this domain, and compares three detectors from three different
        methodological families on the full population of 870,514 apps rather than a
        single detector on a sample.
  \item It tests whether removing the flagged apps improves the removal prediction
        model (Validation 1), under two different training procedures, and reports a negative
        result together with an explanation of why the result is negative.
  \item It validates the flagged apps against subsets of the same dataset whose
        labels VirusTotal and Quark Engine independently confirm (Validation 2), separately
        for each label, and characterizes the direction in which the flagged labels appear to
        be wrong.
\end{enumerate}
 
The remainder of this work is organized as follows. Section~2 reviews the removal prediction
and label noise literature. Section~3 describes the dataset, the three detectors, and the two
validations. Section~4 reports the results of each validation. Section~5 interprets them,
Section~6 states the limitations, and Section~7 concludes and outlines future work.

\section{Related Work}

This research extends the work of \cite{mohsen2022} on their predictive model for detecting whether an app should be removed from Google Play. \cite{mohsen2022} built an XGBoost model predicting Google Play app removal across 870,514 apps, using 47 user-centered features (AUC 0.792) and 37 developer-centered features (AUC 0.762). The user-centered model targets end users and the store evaluating live apps, while the developer-centered model relies only on pre-publication features to aid submission decisions. This work focuses exclusively on the user-centered model.

Moreover, the work of \cite{wang2018} is the closest work aligning with \cite{mohsen2022}, as they investigated why apps get removed from Google Play. Five categories, malicious, fake, spam, apps posing a privacy risk, and apps violating privacy, were the characteristics established to characterize an app as bad. Their work relied on manual analysis, and assumed that the top apps that had the most amount of downloads, reviews and ratings are not considered spam.

\cite{lin2021} did similar work on the iOS App Store, where they built a removal prediction model from metadata features. Like this study, \citet{lin2021} infer removal by checking whether an app has disappeared from the store's listings, though they do so through daily snapshots instead of two observations taken at the start and end of a fixed window, as \cite{mohsen2022} did. Although this tracking is better, it still does not record why the removal occurred, so the concern raised in this work applies there as well.

None of these studies question whether the label itself is correct. That assumption is what this work examines. \cite{frenay2014} define label noise as mislabeled instances in the training data and show that it lowers accuracy, increases model complexity, and increases the number of training samples needed. They categorize noise by what drives the probability of a wrong label. Under noise completely at random (NCAR), any instance is equally likely to be mislabeled regardless of its class or characteristics. Under noise at random (NAR), mislabeling probability differs by class, meaning one class is systematically more affected than the other. Under noise not at random (NNAR), it depends on the instance's own characteristics, so apps sharing the same true class can still differ in how likely they are to carry a wrong label. The noise expected in this dataset is of the third kind, as argued below.
\citet{song2022} survey methods for training deep neural networks under label noise and group 62 of them into five families: robust architecture, robust regularization, robust loss function, loss adjustment, and sample selection. All five intervene in the training procedure
itself. The detectors used here sit outside that taxonomy: they run before training and return a set of suspected instances, which is what makes it possible to inspect the flagged apps directly rather than only through the accuracy of a model trained afterwards.

The dataset of \cite{mohsen2022} may contain noise that is not random because the relationship between an app's market status and the underlying reason for removal could depend on the characteristics of each app. For instance, apps that were withdrawn voluntarily could differ from the ones that were removed due to policy violations. 

\cite{northcutt2021} show that even benchmark datasets (ImageNet, CIFAR, MNIST) all contain pervasive label errors, where at least 3.3\% errors on average across the 10 datasets, e.g. at least 6\% of the ImageNet validation set. The relevance here is that label errors are not exceptional. If curated benchmarks carry them, a dataset labeled by automated observation of a store is unlikely to be free of them. Closer to this domain, \cite{malwhiteout2022} show that Android malware datasets labeled through VirusTotal are themselves prone to mislabeling, and they apply Confident Learning to reduce these errors. Additionally, \cite{brodley1999} were among the first to use classifiers as noise filters for training data. Their approach is similar to the one in this work when Prediction Inconsistency is used: train multiple classifiers, flag instances where classifiers disagree with the label and remove them.

This work applies three label noise detectors: one based on anomalies, using Isolation Forests \cite{liu2008}, one based on neighbors, following the editing rule of \citet{wilson1972} and extended by \citet{tomek1976}, and one based on a model, following the noise filtering approach of \citet{brodley1999}. Both \cite{mohsen2022} and this work use XGBoost \citep{chen2016} as a baseline and in the prediction-based detector.
\section{Methodology}

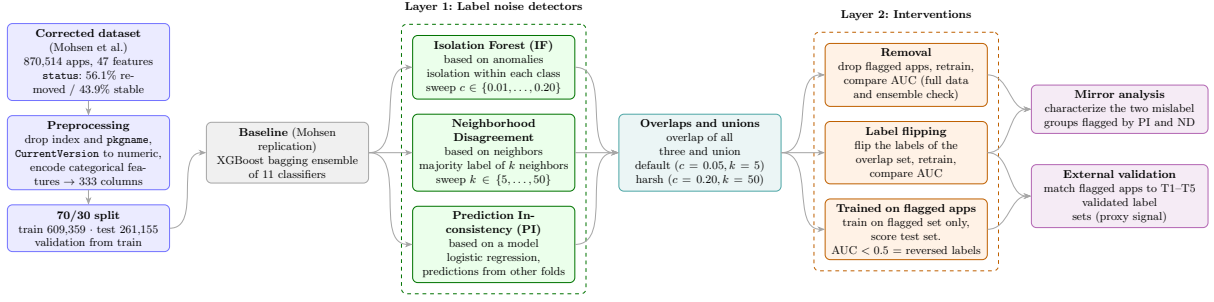
\begin{figure*}[htbp] 
\centering
\resizebox{\textwidth}{!}{%
\begin{tikzpicture}[
  font=\small,
  node distance=4mm and 8mm,
  data/.style   ={rectangle, rounded corners, draw=blue!55, fill=blue!8,
                  line width=0.5pt, align=center, text width=42mm, inner sep=4pt, minimum height=11mm},
  base/.style   ={rectangle, rounded corners, draw=gray!70, fill=gray!12,
                  line width=0.5pt, align=center, text width=42mm, inner sep=4pt, minimum height=11mm},
  det/.style    ={rectangle, rounded corners, draw=green!45!black, fill=green!8,
                  line width=0.5pt, align=center, text width=42mm, inner sep=4pt, minimum height=14mm},
  ov/.style     ={rectangle, rounded corners, draw=teal!70, fill=teal!8,
                  line width=0.5pt, align=center, text width=42mm, inner sep=4pt, minimum height=11mm},
  int/.style    ={rectangle, rounded corners, draw=orange!75!black, fill=orange!10,
                  line width=0.5pt, align=center, text width=42mm, inner sep=4pt, minimum height=14mm},
  ana/.style    ={rectangle, rounded corners, draw=violet!60, fill=violet!8,
                  line width=0.5pt, align=center, text width=46mm, inner sep=4pt, minimum height=12mm},
  arr/.style    ={-{Stealth[length=2mm]}, line width=0.6pt, gray!70},
]

\node[data] (dataset) {\textbf{Corrected dataset} (Mohsen et al.)\\870,514 apps, 47 features\\\texttt{status}: 56.1\% removed / 43.9\% stable};
\node[data, below=of dataset] (prep) {\textbf{Preprocessing}\\drop index and \texttt{pkgname},\\\texttt{CurrentVersion} to numeric,\\encode categorical features $\rightarrow$ 333 columns};
\node[data, below=of prep] (split) {\textbf{70/30 split}\\train 609,359 $\cdot$ test 261,155\\validation from train};

\node[base, right=10mm of prep] (baseline) {\textbf{Baseline} (Mohsen replication)\\XGBoost bagging ensemble\\of 11 classifiers};

\node[det, right=12mm of baseline] (nd) {\textbf{Neighborhood Disagreement}\\based on neighbors\\majority label of $k$ neighbors\\sweep $k \in \{5,\dots,50\}$};
\node[det, above=of nd] (if) {\textbf{Isolation Forest (IF)}\\based on anomalies\\isolation within each class\\sweep $c \in \{0.01,\dots,0.20\}$};
\node[det, below=of nd] (pi) {\textbf{Prediction Inconsistency (PI)}\\based on a model\\logistic regression,\\predictions from other folds};

\node[ov, right=12mm of nd] (overlap) {\textbf{Overlaps and unions}\\overlap of all three and union\\default ($c=0.05, k=5$)\\harsh ($c=0.20, k=50$)};

\node[int, right=12mm of overlap] (flip) {\textbf{Label flipping}\\flip the labels of the\\overlap set, retrain,\\compare AUC};
\node[int, above=of flip] (removal) {\textbf{Removal}\\drop flagged apps, retrain,\\compare AUC (full data\\and ensemble check)};
\node[int, below=of flip] (diag) {\textbf{Trained on flagged apps}\\train on flagged set only,\\score test set.\\AUC $<0.5$ = reversed labels};

\node[ana, right=12mm of flip, yshift=12mm] (mirror) {\textbf{Mirror analysis}\\characterize the two mislabel\\groups flagged by PI and ND};
\node[ana, right=12mm of flip, yshift=-12mm] (val) {\textbf{External validation}\\match flagged apps to T1--T5\\validated label sets (proxy signal)};

\draw[arr] (dataset) -- (prep);
\draw[arr] (prep) -- (split);
\draw[arr] (split.east) to[out=0, in=180] (baseline.west);

\draw[arr] (baseline.east) to[out=0, in=180] (if.west);
\draw[arr] (baseline.east) -- (nd.west);
\draw[arr] (baseline.east) to[out=0, in=180] (pi.west);

\draw[arr] (if.east) to[out=0, in=180] (overlap.west);
\draw[arr] (nd.east) -- (overlap.west);
\draw[arr] (pi.east) to[out=0, in=180] (overlap.west);

\draw[arr] (overlap.east) to[out=0, in=180] (removal.west);
\draw[arr] (overlap.east) -- (flip.west);
\draw[arr] (overlap.east) to[out=0, in=180] (diag.west);

\draw[arr] (removal.east) to[out=0, in=180] (mirror.west);
\draw[arr] (flip.east) to[out=0, in=180] (mirror.west);
\draw[arr] (diag.east) to[out=0, in=180] (val.west);
\draw[arr] (flip.east) to[out=0, in=180] (val.west);

\begin{scope}[on background layer]
  \node[fit=(if)(nd)(pi), draw=green!45!black, dashed, rounded corners, inner sep=3mm, line width=0.4pt] (l1box) {};
  \node[fit=(removal)(flip)(diag), draw=orange!75!black, dashed, rounded corners, inner sep=3mm, line width=0.4pt] (l2box) {};
\end{scope}

\node[font=\small\bfseries, above=2mm of l1box] {Layer 1: Label noise detectors};
\node[font=\small\bfseries, above=2mm of l2box] {Layer 2: Interventions};

\end{tikzpicture}%
}
\caption{Overview of the pipeline. The corrected dataset is preprocessed and split, and the baseline reproduces the ensemble of \citet{mohsen2022}. Layer 1 runs three label noise detectors from different families over the full dataset, whose outputs form the overlap and union sets. Layer 2 intervenes on the flagged apps through removal, label flipping, and the diagnostic trained on flagged apps. The mirror analysis and the external validation then characterize and check the flagged apps.}
\label{fig:pipeline}
\end{figure*}

The full pipeline of this work is summarized in Figure~\ref{fig:pipeline}. The corrected dataset is first preprocessed and split, and a baseline model reproduces the ensemble of \citet{mohsen2022}. Three label noise detectors then form Layer 1 and run over the full dataset, and their outputs are combined into overlap and union sets. Layer 2 acts on the flagged apps through removal, label flipping, and the diagnostic trained on flagged apps. Two further analyses, the mirror analysis and the external validation, characterize the flagged apps and check them against an independent label signal. Each stage is described in the subsections that follow. The work answers two validations: whether removing flagged apps improves the prediction model (Validation 1), and whether flagged apps are under-represented in the label-validated datasets (Validation 2).


\subsection{Terminology}
 
The following terms are used throughout this work.
 
\begin{description}
  \item[Status.] The binary label of the original dataset, where $1$ denotes an app
    that had disappeared from the store at the second observation (\emph{removed}) and $0$
    denotes one that was still present (\emph{stable}).
    \item[Area Under the ROC Curve (AUC).] A measure of how well a binary classifier separates the two classes across all possible decision thresholds. A value of $1.0$ indicates perfect separation and $0.5$ indicates performance no better than random. All reported AUC values are estimated on the test set unless stated otherwise.
  \item[Label noise.] An instance whose status does not match the policy outcome that status
    is being used as a proxy for: an app removed for a reason unrelated to policy
    enforcement, or one that violates policy but has not yet been taken down.
  \item[Flagged.] An app that a detector marks as a suspected mislabel. Flagging is a
    suspicion, not a verdict: the true reason for a removal is not available in this dataset.
  \item[Overlap.] The apps flagged by all three detectors at once. These are treated
    as the strongest mislabeling candidates.
  \item[Union.] The apps flagged by at least one detector.
  \item[Default and harsh settings.] Each detector has a parameter that controls how many
    apps it flags. The default setting is $c = 0.05$ for Isolation Forest and $k = 5$
    for Neighborhood Disagreement, the harsh settings are $c = 0.20$ with $k = 5$, and
    $c = 0.20$ with $k = 50$. Prediction Inconsistency has a single configuration used in all
    three.
  \item[Mirror analysis.] A comparison of the flagged apps against the unflagged
    apps carrying the same label, reported separately for each label, used to
    describe what the suspected mislabels look like.
  \item[Data Partitions.] The four partitions
    defined in Section~\ref{sec:splits}. Removal experiments operate on the training set,
    while the flipping experiment and the diagnostic operate on the trainable partition, so
    the number of flagged apps differs between them.
  \item[Validated datasets $T_1$--$T_5$.] Subsets of the same dataset published by
    \citet{mohsen2026data}, containing the apps whose status agrees with both the
    VirusTotal and the Quark Engine verdict. They are described in
    Section~\ref{sec:validation}. They are not lists of malicious apps: each contains
    both a malicious and a benign side.
\end{description}

\subsection{Dataset}

This work uses the dataset introduced by \citet{mohsen2022}, in the corrected version published by \citet{mohsen2026data}, which contains 870,514 Google Play apps. The main variable is \texttt{status}, which is encoded in binary form. This means that 1 denotes a removed app and 0 denotes a stable one. Moreover, 56.1\% out of the 870,514 apps are labeled as removed. The status label is not a time series. \citet{mohsen2022} observed each app twice: an
app present at the first observation and absent at the second is labeled removed,
and one present at both is labeled stable. Each app therefore contributes a single
row with a single binary outcome, and the features describe the app as it was at the
first observation. Nothing in the data records when between the two observations an
app disappeared, or why. That absence is the reason a status label cannot be read
directly as a record of policy enforcement, and it is what this work investigates.

This work uses the configuration of the dataset centered on the user, which contains the 47 features listed in Appendix~\ref{app:features}. The configuration centered on the
developer contains 37 features and is not analyzed here.

Before modeling, the index column and the \texttt{pkgname} columns are dropped since neither of them carries predictive information. The \texttt{CurrentVersion} column is converted to numeric using pandas' \texttt{to\_numeric} with \texttt{errors="coerce"}. This change turns version strings that are not numeric into missing values. All of the remaining categorical features are encoded into binary indicator columns with \texttt{get\_dummies} and \texttt{drop\_first=True}, which expands the feature space to 333 columns.

\subsection{Train, Validation, and Test Split}
\label{sec:splits}

The data is split twice. First, an outer 70/30 split separates 609,359 apps for training from 261,155 apps for testing. Second, a validation set is drawn from the training portion only, by using a \texttt{test\_size} of 0.4285. This value is taken from \citet{mohsen2022}, who chose it so that the validation set is the same size as the test set. This procedure is described explicitly here because framing it as a ``40/30/30'' split misrepresents how the sets were realistically constructed. This produces four partitions referred to by fixed names throughout: the training set (609,359 apps), the validation set (261,111, drawn once and held fixed), the trainable partition (the 348,248 training apps outside the validation set), and the test set (261,155). The removal from the full data experiments operate on the training set, while the flipping experiment and the diagnostic operate on the trainable partition, so counts of flagged apps differ between the two.

\subsection{Baseline Model}

The baseline model replicates the approach of \citet{mohsen2022}. It is an XGBoost \citep{chen2016} bagging ensemble that is composed of 11 classifiers. Each classifier is trained with randomized \texttt{max\_depth} and \texttt{n\_estimators} values, and each is trained on a balanced subset of the data to address the class imbalance between removed and stable apps. This baseline serves as a point of comparison for every intervention that is later applied in this work. For the validation experiments, the same ensemble is also trained on the entire training set, with scale\_pos\_weight handling the class imbalance in place of the balanced subsets.

\subsection{Label Noise Detectors}

Three label noise detectors are applied to the dataset, each of which is drawn from a different detection approach. These approaches are detection based on anomalies, detection based on neighbors, and detection based on a model. This forms the Layer 1 of this work's pipeline, thus producing the sets of flagged apps on which Layer 2 later intervenes. All three detectors are run on the full dataset rather than on a sample, so that the overlap between them is computed over every app and not over a subset that each detector happened to see.

\subsubsection{Isolation Forest}

Isolation Forest \citep{liu2008} detects anomalies by repeatedly splitting the data along
randomly chosen features at randomly chosen values, and recording how many splits it takes to
separate each point from the rest. Points that sit in sparse regions of the feature space are
separated after few splits, and this short average path length is what marks them as
anomalous. Unlike distance- or density-based detectors, the method never computes how far
apart two apps are, which makes it inexpensive on a dataset of this size. Here it is
fitted separately within each label class, so an app is flagged when it is unusual
relative to other apps carrying the same label, rather than unusual overall.

The proportion of points the method returns as anomalies is set by a parameter called the
contamination factor, which has to be supplied in advance. Choosing it is a known difficulty.
\citet{perini2022} observe that the contamination factor is generally unknown in practice and
can only be approximated by labeling part of the data, and \citet{campos2016} add that even
when such labels exist, how outlier detectors respond to their parameter settings remains
poorly understood. Both obstacles apply with more force here, because the labels of this
dataset are the object of study rather than a trusted reference, so the usual fallback of
estimating the contamination factor from known-good labels is unavailable. The parameter is
therefore reported across five values: $c \in \{0.01, 0.05, 0.10, 0.15, 0.20\}$, ranging from conservative through aggressive settings. It is expected that the number of flagged instances increases as the value rises. The contamination factor is reported as a full span because it is intended to show how much the detector's output depends on this setting instead of relying only on a single value that could not be justified. When a single setting is needed for the overlaps and unions defined later, $c = 0.05$ is used for the default configuration and $c = 0.20$ for the harsh one.

\subsubsection{Neighborhood Disagreement}

The Neighborhood Disagreement (ND) detector flags an app when the majority of its $k$ nearest neighbors (kNN) have the opposite label. This follows the editing principle introduced by \citet{wilson1972}, who used a kNN majority vote to identify instances that a classifier disagrees with. This rule was further extended by \citet{tomek1976} by repeating the edit for increasing values of $k$. The choice of $k$ therefore changes what is flagged as noise. This is the same difficulty the contamination factor raises for Isolation Forest, and it is handled the same way here, by reporting a range of values rather than a single one. Therefore, this work uses $k \in \{5, 10, 20, 50\}$ to check whether the flags stay stable as the number of neighbors changes. Across this range, the flag rate stays close to 27\% for every value of $k$. This means the flagged set stays around the same size regardless of which $k$ is picked, so the sweep does not depend on one arbitrary setting.

\subsubsection{Prediction Inconsistency}

The Prediction Inconsistency (PI) detector flags an app when a classifier trained on
the dataset predicts a different label from the one the app actually carries. The
reasoning is that a classifier learns the patterns that hold across the data as a whole, so an
app it confidently places in the opposite class is one whose features do not match its
label. This use of a classifier as a filter for suspect training data goes back to
\citet{brodley1999}.

The classifier used here is logistic regression rather than XGBoost. This is deliberate: the
model whose performance is later measured is an XGBoost model, and using XGBoost to decide
which apps that same model should be trained on would let the detector remove exactly
the apps the predictor finds difficult, which would make any improvement circular.

A classifier cannot be used to judge an app it was trained on, because it has already seen that app's label and will tend to agree with it. To avoid this while still
covering every app, predictions are generated out of fold: the data is divided into
five stratified folds, and each app is scored by a model trained on the other four,
so no app is ever judged by a model that saw it. This is the same separation of
training and scoring that underlies stacked generalization \citep{wolpert1992}. The result is
that PI covers the full population of 870,514 apps, which makes its flagged set
directly comparable with those of Isolation Forest and Neighborhood Disagreement.

\subsection{Overlaps and Unions}

Because the detectors operate on different principles, their outputs are compared through overlap and union sets, and each detector is also assessed individually. This work defines an overlap (the intersection) where an app is flagged by Isolation Forest, Neighborhood Disagreement and Prediction Inconsistency simultaneously. Moreover, a union occurs when an app is flagged by at least one detector. These are computed at three settings: default (IF $c =0.05$ and ND $k = 5$), harsh k = 5 (IF $c = 0.20$, ND $k = 5$), and harsh k = 50 ($c = 0.20$ and $k = 50$). Prediction Inconsistency has a single configuration shared across all three.

\subsection{Interventions}

Once the apps are flagged, this work tests whether flagging affects the XGBoost's model performance: First, by removing or flipping the labels of the flagged apps and secondly, by checking whether the flagged apps behave as if their labels were reversed. 

\subsubsection{Removal}

The first intervention consists of removing the flagged apps from the training set and retrains the baseline model. This AUC result is then compared against the original baseline. Note that only flagged apps present in the training set are removed, which means that the test set is never altered. Therefore, any change in performance reflects the effect of cleaning the training data on the testing set. 

\subsubsection{Label Flipping}

The second intervention consists of flipping the labels of the apps only in the overlap set under the assumption that apps flagged by all of the detectors are the strongest mislabeling candidates. The model is retrained on this relabeled training set afterwards. 

\subsubsection{Diagnostic Trained on Flagged Apps}

The third intervention is evaluative, where a model is trained only using the flagged sets of the apps and then tested on the test set. 

If the resulting AUC is below 0.5, then this indicates that the relationship between features and labels is reversed relative to the test distribution. This is evidence that there is a systematic label inconsistency, but because there are no verified removal reasons shown in the data, then the diagnostic cannot confirm whether every flagged app is mislabeled.

\subsubsection{Validation Set Control}

The validation set is used to compute the AUC-val, which is drawn only once and held fixed throughout every subsequent experiment. This is explicitly stated since methodologically, if the validation set would be drawn again or cleaned like the training data, then any improvement in AUC-val could show an easier validation set, but it does not necessarily mean that the model's performance improved. Since the detector's outputs were created using the full population, this analysis describes the available dataset instead of testing how these detectors will perform with unseen datasets. Therefore, this intervention results should be interpreted as internal analysis of this dataset. This fixed validation set applies to the intervention experiments. The replication reported in Section 4.1 follows the original procedure of \cite{mohsen2022}, in which the validation sample is redrawn for each run, and its validation AUC is therefore reported only for replication purposes.

\subsection{Validation}
\label{sec:validation}

To check the flagged apps against an independent signal, this work matches them by pkgname against the validated datasets of \citet{mohsen2026data}. Each validated set Ti contains the apps whose status label agrees with both external tools. Its malicious side holds removed apps detected by at least i VirusTotal engines and rated at least Moderate Risk by Quark Engine, and its benign side holds stable apps with fewer than i detections and a Low Risk rating. Every set therefore contains both labels, and the three-way agreement gives these labels higher confidence than the rest of the dataset. The two sides move in opposite directions as i grows: confidence in the removed label increases from T1 (62,093 removed, 18,275 stable) to T5 (19,027 removed, 19,275 stable), while the benign criterion loosens.

For each detector set, the share of flagged apps is computed inside the original dataset and inside the matching side of each validated set. If the detectors flag mislabeled apps, they should be under-represented where labels are confirmed. Two qualifications apply. Absence from every validated set is ambiguous between never scanned and scanned but disagreeing. In addition, the scanned apps skew much older than the store average, so rates are also reported after standardizing on the age distribution of each validated cohort, using deciles of days since last update.

The inconsistency can run in either direction. An app may be flagged because it
carries the removed label while looking like an app that would have been allowed to
stay, or because it carries the stable label while looking like one that should have been
taken down. The analysis therefore reports the two labels separately throughout, rather than treating the flagged apps as a single group.

\subsection{Evaluation Metrics}

The principal evaluation metric used throughout this work is AUC. This metric is computed on the test set for every run of the baseline and intervention. Subsequently, precision, recall and F1 are reported alongside AUC to provide a wider scope of how each intervention affects the model's classification behavior. For instance, a decrease in precision but an increase in recall means that an intervention may be prioritizing to catch all true app removals even if it results in more false alarms.

The importance of each feature in the baseline model is also recorded, averaged across the classifiers of the ensemble. For the twenty most important features, the mean value is computed separately for the apps labeled as removed and for those labeled as stable, which shows the direction in which each feature separates the two classes.
\section{Results}

\subsection{Baseline}
\label{sec:baseline}

The baseline reproduces the model of \citet{mohsen2022} on their dataset. Trained on a balanced subset of 100,000 apps, it reaches an AUC of 0.7937 on the test set, with an AUC of 0.7946 on the validation set. Precision, recall, and F1 on the test set are 0.7665, 0.7064, and 0.7352. This is close enough to the 0.792 that \citet{mohsen2022} report, so the replication works. Every removal and flipping result below is compared against this same model. All removal comparisons in this work are made against models trained on the full training population (Section 4.3), the ensemble replication in this section serves only to reproduce \citet{mohsen2022}.

\subsection{Detector Agreement}

The three detectors flag very different amounts of the data. At its default contamination of 0.05, Isolation Forest flags 43,527 apps, which is close to 5\% of the dataset by construction. Neighborhood Disagreement at $k=5$ flags 232,125 apps, and Prediction Inconsistency flags 264,694, so these two detectors each mark roughly a quarter to a third of the dataset. Prediction Inconsistency skews toward stable apps, since 44.2\% of its flagged set carries the removed label against the 56.1\% base rate.

\begin{figure*}[htbp]
\centering
\includegraphics[width=\textwidth]{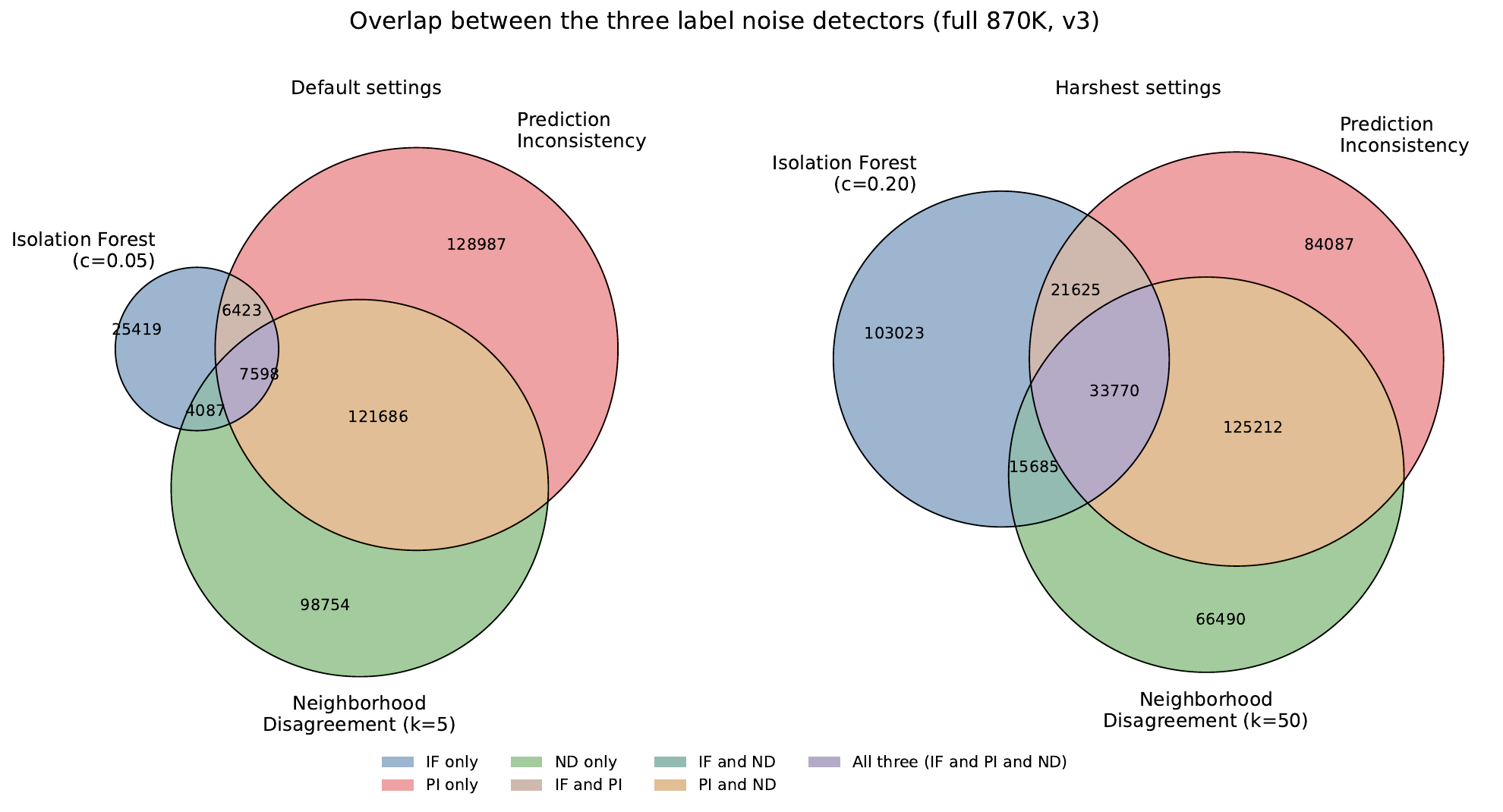}
\caption{Overlap between the three detectors on the full dataset, at default settings on the left and harsh settings on the right. All three agree on 7,598 apps at the default setting, a small fraction of the 392,954 in the union.}
\label{fig:venn}
\end{figure*}

Where the detectors agree is far smaller. Only 7,598 apps are flagged by all three at once, which is under 1\% of the data. The union of the three flags 392,954 apps, which is close to 45\% of the dataset. Figure~\ref{fig:venn} shows the overlap structure. The gap between the union and the overlap of all three detectors is the first indication that the detectors respond to different properties of the data, and that their point of agreement is narrow.

\subsection{Validation 1: Removing Flagged Apps}

Removing flagged apps from the training set does not improve the model in any setting. This experiment trains a single model on the entire cleaned training set, so its baseline of 0.7927 comes from that procedure and is separate from the ensemble reported in Section~\ref{sec:baseline}. Table~\ref{tab:removal_full} in the appendix reports the results against this baseline of 0.7927. The smallest change comes from removing the overlap of all three detectors, which moves the AUC by $-0.0003$ to 0.7924. Every other removal lowers the AUC, and the size of the drop tracks the size of the flagged set. Removing the default set of Isolation Forest costs $-0.0018$, the default set of Neighborhood Disagreement costs $-0.0048$, and the set of Prediction Inconsistency costs $-0.0232$, which is the largest drop for a single detector. Removing the unions costs around $-0.022$ across settings. Pushing the detectors to their harsh settings makes matters worse. Isolation Forest at $c=0.20$ falls to 0.7825 and Neighborhood Disagreement at $k=50$ falls to 0.7796.

\begin{table*}[htbp]
\centering
\caption{Validation 1. The bagging ensemble of 11 XGBoost classifiers is trained on the full training population, with \texttt{scale\_pos\_weight} replacing the balanced subsets. The flagged apps of each detector at default settings are removed from the training set, the model is retrained, and the result is compared with the baseline on the untouched test set. No removal improves on the baseline.}
\label{tab:validation1}
\resizebox{\textwidth}{!}{%
\begin{tabular}{lrrrrrrr}
\toprule
Model & Removed & Train $n$ & AUC & $\Delta$AUC & P & R & F1 \\
\midrule
Baseline (no removal) & 0 & 609,359 & 0.7908 & --- & 0.7642 & 0.7035 & 0.7326 \\
Isolation Forest ($c=0.05$) & 30,430 & 578,929 & 0.7882 & $-0.0026$ & 0.7643 & 0.7015 & 0.7316 \\
Neighborhood Disagreement ($k=5$) & 162,690 & 446,669 & 0.7852 & $-0.0056$ & 0.7556 & 0.7183 & 0.7365 \\
Prediction Inconsistency & 185,451 & 423,908 & 0.7681 & $-0.0227$ & 0.7251 & 0.7503 & 0.7375 \\
Overlap (all three) & 5,288 & 604,071 & 0.7899 & $-0.0009$ & 0.7637 & 0.7032 & 0.7322 \\
Union (all three) & 275,254 & 334,105 & 0.7690 & $-0.0218$ & 0.7258 & 0.7520 & 0.7387 \\
\bottomrule
\end{tabular}%
}
\end{table*}

Under the ensemble trained on the full population (Table 1), the pattern is identical to the single-model results in Table 5: the overlap costs $-0.0009$, Isolation Forest $-0.0026$, Neighborhood Disagreement $-0.0056$, and Prediction Inconsistency and the union around $-0.022$. The conclusion therefore does not depend on the training procedure.

\begin{figure*}[htbp]
\centering
\includegraphics[width=0.85\textwidth]{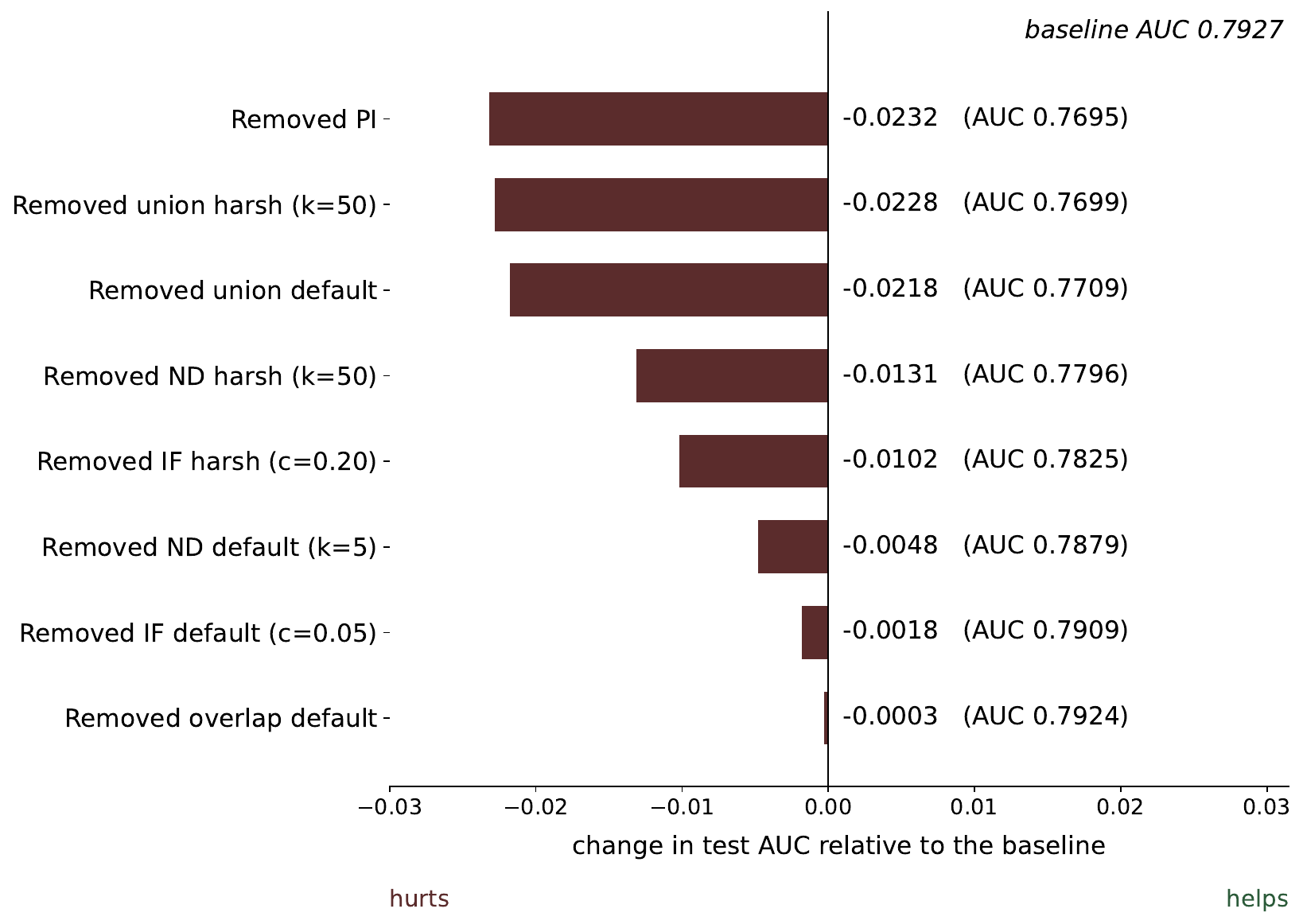}
\caption{Change in test AUC when each flagged set is removed from the training data, measured against the baseline. Every setting lowers the AUC, and the size of the drop grows with the size of the flagged set.}
\label{fig:removal}
\end{figure*}

Figure~\ref{fig:removal} shows the same pattern, where the interventions that remove the most data cause the largest losses, and none of them lift the performance above the baseline. Precision and recall shift in opposite directions as more apps are dropped, with recall rising and precision falling, which explains why F1 stays close to flat even as AUC decreases. The practical reading is that the flagged apps, taken as a group, carry useful signal for the removal task, so discarding them removes information the model was using.

The pattern also appears when removal is repeated under the ensemble of the baseline, where across every training size, removing any flagged set leaves the AUC at or below the baseline, with Prediction Inconsistency and the unions again causing the largest drops. Table~\ref{tab:removal_ensemble} in the appendix reports these results at the 100,000 size. Because the two procedures reach the same conclusion, the effect of removal does not depend on whether the model is trained on the full pool or on balanced subsets.

\subsection{Label Flipping}

Flipping the labels of the apps the detectors agree on produces no improvement either. Table~\ref{tab:flip} in the appendix reports the flipping results at the 100,000 training size against a baseline of 0.7941. By flipping the overlap of all three detectors, the AUC is moved by $-0.0006$ to 0.7935. Flipping the harsh overlap at $k=5$ moves it by $+0.0001$ to 0.7942, which is within noise of the baseline. The harsh overlap at $k=50$ also costs the AUC $-0.0055$. Therefore, none of these changes is sufficiently large to see this result as significant. An earlier version of this experiment, which was run before the validation set was held fixed, had shown a small AUC increase from flipping. However, that gain did not survive the leakage correction, and the corrected results show that flipping the label has no effect. 

\subsection{Diagnostic Trained on Flagged Apps}

As stated previously, the removal and flipping results show that cleaning the flagged apps does not help the model. They do not show whether those apps are actually mislabeled. However, the diagnostic answers that question. A model is trained using only a flagged set, and then it is scored on the test set. For reference, a model trained on the full trainable pool reaches an AUC of 0.8029 under this setup. A flagged set whose relationship between features and labels matches the original test distribution produces a model with an AUC above 0.5. Conversely, a model that is trained on a flagged set that has a reversed relationship between the features and labels could produce an AUC that is below 0.5 when it is evaluated on the test set.

\begin{table}[htbp]
\centering
\caption{Diagnostic. A classifier is trained only on each flagged set and scored on the test set. An AUC that scores below 0.5 means that a model trained on the selected set is predicting in the opposite direction relative to the original test distribution. This is consistent with a systematic label inconsistency, yet it does not confirm that every flagged label is incorrect. PI, ND, and all overlap sets are strongly reversed, while the Isolation Forest sets stay above 0.5, which points to feature anomalies rather than reversed relationships. \%Rem is the share of apps labeled as removed in each set, and the reversed sets skew stable. Rows are ordered by test AUC.}
\label{tab:trainedon}
\resizebox{\columnwidth}{!}{%
\begin{tabular}{lrrrrrr}
\toprule
Set & $n$ & AUC-test & P & R & F1 & \%Rem \\
\midrule
Overlap harsh ($k=5$) & 11,120 & 0.2417 & 0.3385 & 0.2440 & 0.2836 & 38.9 \\
Overlap harsh ($k=50$) & 13,395 & 0.2426 & 0.3422 & 0.2509 & 0.2895 & 40.4 \\
PI & 106,076 & 0.2431 & 0.3395 & 0.2435 & 0.2836 & 44.2 \\
Overlap default & 3,021 & 0.2518 & 0.3729 & 0.3094 & 0.3381 & 43.3 \\
ND default ($k=5$) & 93,325 & 0.3642 & 0.4521 & 0.3127 & 0.3697 & 46.4 \\
Union default & 157,746 & 0.3861 & 0.4591 & 0.3478 & 0.3958 & 48.2 \\
Union harsh ($k=50$) & 180,409 & 0.4772 & 0.5362 & 0.4389 & 0.4827 & 50.0 \\
Union harsh ($k=5$) & 188,099 & 0.5328 & 0.5717 & 0.4840 & 0.5242 & 50.8 \\
IF default ($c=0.05$) & 17,392 & 0.6039 & 0.6427 & 0.5111 & 0.5694 & 56.4 \\
IF harsh ($c=0.20$) & 69,529 & 0.6801 & 0.6937 & 0.5746 & 0.6286 & 56.1 \\
\midrule
Baseline (full trainable) & 348,248 & 0.8029 & 0.7390 & 0.7908 & 0.7640 & 56.1 \\
\bottomrule
\end{tabular}}
\end{table}

Table~\ref{tab:trainedon} reports these results. When the model is trained on the overlap of all three detectors with the harsh $k=5$ setting, it produces an AUC of 0.2417. Then, the default overlap results in 0.2518, 0.2431 for Prediction Inconsistency, and 0.3642 for Neighborhood Disagreement. 

All of these results fall under 0.5, which means the model predicts the opposite direction relative to the test set. This means that the relationship between labels and features shows a pattern of a systematic label inconsistency, but it does not confirm that all flagged apps are incorrect. 

On the other hand, training the model on the apps flagged by Isolation Forest, gives AUCs of 0.6039 at the default and 0.6801 at the harsh setting, and both are above 0.5. This suggests that Isolation Forest identifies apps that are unusual in terms of features, but it is different from label inconsistency because an unusual app could still have a correct observed label. Therefore, a model trained on these apps continues to predict the data correctly.

\begin{figure*}[t]
\centering
\includegraphics[width=0.85\textwidth]{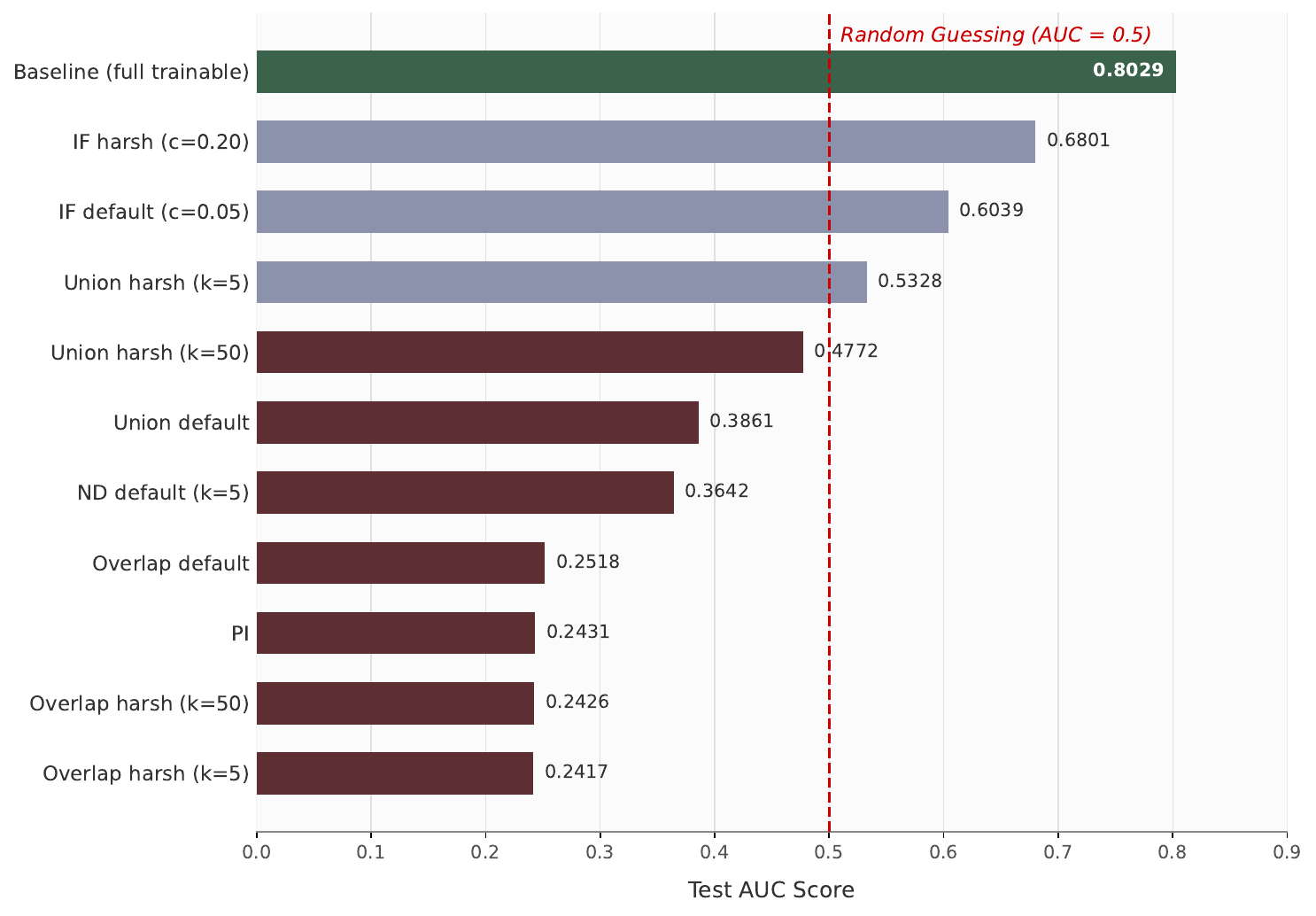}
\caption{Test AUC of a model trained using only each flagged set. Values below the 0.5 chance line indicate that the labels of the set run opposite to its features. The overlap, Neighborhood Disagreement, and Prediction Inconsistency are strongly reversed, while Isolation Forest stays above the line.}
\label{fig:trainedon}
\end{figure*}

 The unions fall between the two behaviors, where the default union results in an AUC of 0.3861 and the harsh $k=5$ union at 0.5328. This is because they mix the reversed apps from Neighborhood Disagreement and Prediction Inconsistency with the anomalous ones from Isolation Forest. Figure~\ref{fig:trainedon} makes the split visible. The overlap, Neighborhood Disagreement, and Prediction Inconsistency sit well below the chance line, and Isolation Forest sits above it. Therefore, mixing the two kinds of flagged apps pulls the AUC toward 0.5.

\subsection{Feature Directions}

Figure~\ref{fig:importance} in the appendix reports the twenty most important features of the baseline model, and Table~\ref{tab:feature_directions} reports the mean value of each one computed separately for each label. Most directions are expected. Apps labeled as removed carry an unrated content rating far more often, at 0.21 against 0.03, they have gone longer without an update, with a mean of 614 days against 371, and they have fewer downloads.

Two directions are counterintuitive. The presence of a link to a privacy policy is the second most important feature in the model, and it is higher in the apps labeled as removed, at 0.76 against 0.55. The presence of a field for a developer website is the fourth most important feature, and it is also higher in the apps labeled as removed, at 0.43 against 0.33. Apps that declare more complete metadata are therefore removed more often. The spam indicator is the third most important feature and points the same way, at 0.75 against 0.59. This is consistent with the explanation that apps engaged in spam fill these fields to satisfy the review process, although the metadata alone cannot confirm it.

\subsection{Characterizing the Mislabels}

The mirror analysis examines what the mislabeled apps look like in terms of their metadata. It compares the apps flagged as suspected mislabels against the normal apps carrying the same label, where normal means flagged by neither detector. Table~\ref{tab:mirror} reports the two groups, and the result is that every feature examined flips sign between them.

\begin{table}[htbp]
\centering
\caption{Mirror analysis of the apps flagged by both PI and ND ($n=129{,}284$). Each suspicious half is compared with the normal apps of the same label, where normal means flagged by neither detector. The two columns give the difference between the suspicious and the normal group. The two differences have opposite signs among every feature, where the suspicious apps labeled as stable look abandoned. These apps have longer gaps since update, less downloads, shorter text and more spam. On the other hand, the suspicious apps labeled as removed look healthy. This means that the two groups are mirror images, so the label errors are in both directions.}
\label{tab:mirror}
\resizebox{\columnwidth}{!}{%
\begin{tabular}{lrr}
\toprule
& Suspicious stable & Suspicious removed \\
Feature & $-$ normal stable & $-$ normal removed \\
\midrule
Days since last update & $+322.8$ & $-391.8$ \\
Max downloads (log) & $-2.44$ & $+2.19$ \\
Description length & $-11.45$ & $+8.68$ \\
What's new length & $-1.04$ & $+1.11$ \\
Title length & $-1.30$ & $+0.42$ \\
Is spamming & $+0.345$ & $-0.330$ \\
Privacy policy link & $+0.455$ & $-0.410$ \\
Developer website & $+0.249$ & $-0.235$ \\
Reviews average & $-0.48$ & $+0.76$ \\
Last updated (year) & $-0.96$ & $+1.13$ \\
\bottomrule
\end{tabular}}
\end{table}

The apps labeled as stable but flagged as suspicious resemble apps that should have been removed. They have gone much longer without an update, where they have a mean of 580 days since the last update against 257 for normal stable apps. They have fewer downloads, shorter descriptions, and higher rates of the spam indicator, the presence of a link to a privacy policy, and the presence of a field for a developer website. The apps labeled as removed but flagged as suspect show the mirror image. They have been updated more recently, carry more downloads and longer descriptions, and show lower rates of the spam indicator and of these two metadata fields. These apps resemble healthy stable apps that were removed for reasons the metadata does not explain.

This pattern is consistent with the feature directions that are reported in Table~\ref{tab:feature_directions} in the appendix and with the underlying behavior observed throughout the dataset. The presence of filled metadata fields is associated with removal, most likely because spam apps add them to pass review. The mirror analysis shows the detectors separate the two labels along exactly the axes that define removal risk, and that the flagged apps are on the wrong side of those axes for the label they carry.

\subsection{Validation 2: Representation in the Validated Datasets}

Table~\ref{tab:flag-rates} reports the results per label. On the removed side, every detector set is under-represented among confirmed removals, and the depletion deepens as the confirmation threshold rises: the default overlap falls from 0.81 times the expected rate at T1 to 0.43 at T5 (0.42 after age adjustment). The removals the detectors question are precisely those without independent malware evidence. On the stable side, the flagged apps appear more often than average in the benign-confirmed sets, up to 3.5 times for the overlap, but this excess is a coverage effect: the benign cohorts average roughly 1,000 days since their last update against 371 for stable apps overall, and after standardizing on that age distribution the ratios fall to between 0.99 and 1.10 for every set. An aggregate reading of these matches as a malware signal would therefore be misleading: at T5, 756 of the overlap's 811 matches fall on the benign side, so most matched apps carry a confirmed benign verdict rather than a malware one.
\section{Discussion}

The results answer the research question in two parts, and the two parts point in different
directions. The first part asked whether the detectors can identify apps whose status
label is inconsistent with policy enforcement. They can: the apps the three detectors
agree on behave as a systematically mislabeled group, in both directions, and they are
depleted among the removals that VirusTotal and Quark Engine independently confirm. The
second part asked whether removing those apps improves the removal prediction model.
It does not: no removal setting, under either training procedure, improves on its baseline.
Together the two findings locate the label noise without offering a way to clean it.

\subsection{The flagged apps show systematic label inconsistency}

The strongest evidence of systematic label inconsistency comes from the diagnostic trained on the flagged apps. The default overlap had 7,598 apps across the complete dataset of which 3,021 of these were in the trainable partition. A model that was trained on only these 3,021 apps had an AUC of 0.2518 on the testing partition, while a model trained on apps flagged by Prediction Inconsistency got an AUC of 0.2431. Both of these values are below the random chance threshold of 0.5, therefore this indicates that the feature label relationships learned from these subsets are reversed relative to the test partition. Consequently, this pattern reflects a systematic label inconsistency, yet it still does not prove that every flagged app is mislabeled. 

Isolation Forest fails to show the same behavior, thus confirming this reading. Its flagged sets score 0.6039 and 0.6801, and both values are above random chance. This detector looks for points that are hard to place in the feature space, and being hard to place is a different property from carrying the wrong label. The contrast between the two behaviors is useful because it shows that the diagnostic measures label correctness instead of flagging whatever is strange.

The mirror analysis explains what the mislabeling looks like. The apps labeled as stable but flagged as suspicious have gone 580 days on average without an update, against 257 days for ordinary stable apps, and they have higher rates of the spam indicator and of the filled metadata fields. They look like apps that should have been removed. The apps labeled as removed but flagged as suspicious show the opposite on every feature examined, with more downloads, longer descriptions, and more recent updates. They look like healthy apps that were removed for reasons the metadata does not record. This pattern is compatible with the study of \citet{wang2018}, which observed that apps could disappear from the store for unrelated reasons to policy violations, which include voluntary withdrawals. Therefore, it is true that the ``removed'' status is an accurate record of an app's market status, but it does not represent the policy outcome that the prediction model is intended to predict.  

This is what \citet{wang2018} anticipated when they observed that developers withdraw apps voluntarily, and it is consistent with the category of noise described by \citet{frenay2014} in which the probability of a wrong label depends on the instance itself.

In terms of external validation, the direction of the effect depends on the label. Removed apps flagged by the detectors are strongly under-represented among removals whose malicious status VirusTotal and Quark Engine both confirm, and the depletion deepens with the confirmation threshold, reaching 0.42 times the expected rate for the overlap after age adjustment. This is consistent with the mirror analysis and with the voluntary withdrawals described by \citet{wang2018}: the questioned removals are the ones without malware evidence. The stable side shows no informative signal once scan coverage is controlled, and benign agreement is weak confirmation of a stable label in any case, because the suspected noise mechanism for stable apps, unremoved spam or abandonment, leaves no malware signature.

\subsection{Why cleaning does not improve the model}

The obvious expectation is that removing mislabeled training data should produce a better model, however it does not. Every removal setting leaves the AUC at or below its baseline (0.7927 for the single full-data model and 0.7908 for the full-population ensemble), and the effect holds whether the model is a single classifier trained on the full dataset or the ensemble of \citet{mohsen2022} trained on balanced subsets. Flipping the labels of the overlap is equally flat, and the largest movement in either direction is a loss of 0.0055.

Two factors explain these findings, where the first one is size. The overlap of all three detectors contains 7,598 apps out of 870,514, which is under 1\% of the data. Even if every one of those labels is wrong, correcting fewer than one app in a hundred cannot be expected to shift an aggregate metric computed over 261,155 test apps. The diagnostic detects the reversal because it isolates those apps and looks at nothing else.

The second factor is that removing flagged apps also removes useful information. The performance drop relates to the size of the removed group. Removing the overlap group causes an AUC drop of 0.0003. Removing the Prediction Inconsistency group causes a drop of 0.0232. Removing the union groups causes a drop of 0.0228. The Prediction Inconsistency group contains 264,694 apps, which is around one third of the dataset. These apps are the difficult cases that the model uses to define a decision boundary. Removing these cases hurts the model more than it helps by removing the noise.

A separate problem is that cleaning the training set does not address the noise in the test set. The training set and the test set contain the same errors because they follow the same collection process. A model trained on corrected labels is still tested against incorrect labels, so improvements in the model still appear as errors during the evaluation. The evaluation compares the model against a noisy standard. Therefore, this prevents the results from showing the full benefit of cleaning the labels.

\subsection{What this means for removal prediction}

Label noise is small enough that a model trained on the uncleaned data is not badly damaged by it. The AUC of 0.792 reported by \citet{mohsen2022} holds up under this scrutiny, and this work reproduces it at 0.7937. The contribution here is a characterization of which apps the labels get wrong and in which direction, and that characterization is usable in a way an aggregate score is not. A moderation team has something concrete to act on: 7,598 apps that three independent methods flagged as suspected mislabeled, and whose removals mostly lack independent malware confirmation.

\section{Limitations}

The most important limitation is that the interventions this work tests do not improve the model, so the second part of the research question receives a negative answer. This negative result does not mean the detectors identified the wrong apps. The diagnostic shows that the flagged sets have feature, label relationships reversed relative to the test distribution, which is evidence that they are systematically inconsistent. What this work cannot do is recommend removal or label flipping as a way to improve removal prediction.

No statistical significance tests were conducted on the AUC differences between the baseline and the cleaned variants. The differences are small in absolute terms, and without confidence intervals or hypothesis tests it is not possible to determine whether they reflect a significant effect or sampling variation in the test set.

The test set is never cleaned, but this is deliberate because cleaning it would make the evaluation circular. However, it means every AUC reported here is measured against labels that are known to contain errors. If the test set carries a similar proportion of reversed labels to the training set, then a perfectly corrected model would still be penalized for disagreeing with them.

The external validation is a proxy and cannot serve as ground truth. \citet{malwhiteout2022} show that VirusTotal verdicts contain errors of their own, so even the three-way agreement behind the validated sets is not infallible, and this analysis cannot determine the actual reason why each app disappeared from the store, so it cannot distinguish conclusively between a voluntary withdrawal, a policy-related removal, or other causes. The validation also depends on which apps were scanned. Absence from every validated set cannot distinguish an unscanned app from one whose label the tools contradicted, and the scanned cohorts skew old, with benign cohorts averaging roughly 1,000 days since the last update against 371 for stable apps overall. The per-label rates are therefore reported with an age-standardized version, although residual coverage effects along other dimensions cannot be excluded. Finally, agreement between the label and the verdicts is asymmetric evidence: it is strong confirmation for a removed label and weak confirmation for a stable one, since policy violations such as spam leave no malware signature, and an app can be removed for a violation that no antivirus engine would flag.

Additionally, the detectors carry assumptions already. Isolation Forest requires a contamination value, which cannot be estimated without trustworthy labels, and this dataset does not provide them. Reporting the detector across a range of contamination rates shows how much its output depends on the setting, but it does not identify which setting is correct. Then, Neighborhood Disagreement depends on a distance metric over the encoded categorical features, and distances between categorical levels are not significant in the way they are for continuous features. Finally, Prediction Inconsistency uses logistic regression, so anything a linear model cannot represent will be flagged as inconsistent even when the label is correct. This last point may explain why Prediction Inconsistency flags so much of the dataset and why removing its flagged set costs the most. Because the detectors were applied to the full population rather than fitted on one
part of the data and tested on another, these results describe this dataset only, and say nothing about how the same methods would behave on data they have not seen.

Lastly, the analysis uses metadata only so the features describe what an app declares about itself, and none of them describe what its code does. Two apps with identical metadata can behave very differently once they are installed, and no method operating on this feature set can distinguish them. The three detectors were chosen to represent three different families of method, not as an exhaustive survey, and they were applied to a single dataset. Nothing here establishes how these detectors would behave on other tabular datasets, or that a fourth detector would not flag a different set of apps.
\section{Conclusion and Future Work}

This work asked whether label noise detectors can identify apps in the Google Play removal dataset of \citet{mohsen2022} whose status label is inconsistent with policy enforcement, and whether removing those apps improves the removal prediction model. In this study, label noise is defined as an inconsistency between the observed market status and its interpretation as an indicator that an app was removed due to a policy violation. 

On the first part, the answer is yes because three detectors for systematic label inconsistency that were drawn from different methodological families were applied to the Google Play dataset of \citet{mohsen2022}. 

The default overlap consisted of 7,598 apps across the complete dataset where the model trained on the 3,021 apps in the trainable partition achieved an AUC of 0.2518 on the test set. This means that it is much lower than the random chance threshold of 0.5. This indicates that the features and labels relate to each other in the opposite way of the labels in the test set, which is a signal of systematic label inconsistency. The mirror analysis shows that the reversal runs in both directions, where abandoned apps that resemble spam carry the stable label, and healthy apps carry the removed label. Matched against the validated datasets, the flagged removals are depleted among confirmed removals, at 0.43 times the expected rate at the strictest threshold, which is external evidence consistent with the mislabel reading, while the stable-side excess reduces to scan-coverage effects. Since the actual removal reasons are unknown, these findings should be treated as strong label noise candidates rather than confirmed errors.

On the second part, the answer is no, since no removal setting improves on the baselines (0.7927 for the single full-data model and 0.7908 for the full-population ensemble), and no flipping setting moves the model. The cluster of label noise candidates covers less than 1\% of the dataset, which is too small to shift an aggregate metric. Moreover, the detectors that flag enough apps to matter also flag the difficult cases that the model needs.

Taken together, the two answers show what detector agreement is good for in this setting. Its value lies in identifying and characterizing the label noise candidates, not in cleaning the training set, which does not improve the model.

The next step follows from the ceiling problem, where a corrected test set, built by reviewing manually a sample of the flagged apps, would allow a cleaned model to be evaluated against labels that are known to be right. Without this, no cleaning intervention can be measured fairly, and the negative result reported here cannot be distinguished from an artifact of the evaluation.

The mirror analysis suggests that relabeling is more appropriate than removal. The two suspicious groups are separable by their label, and each group looks like the opposite class on every feature examined. Assigning them the label their features indicate, instead of discarding them, would keep the training signal and correct the error. This work tested flipping only on the overlap of all three detectors, and a targeted relabeling informed by the mirror directions may behave differently.

Confident Learning, as introduced by \citet{northcutt2021cl} and applied by \citet{malwhiteout2022} to VirusTotal labels, offers a principled way to estimate the joint distribution between observed and true labels. It would provide a probability of mislabeling for each app instead of a binary flag. This suits a dataset where the noise is concentrated and depends on the instance.

The per-label validation could be strengthened with the complement of the validated datasets: the apps whose label the VirusTotal and Quark Engine verdicts contradict. That set would provide direct external mislabel candidates to intersect with the detector flags. The coverage adjustment could also be extended beyond app age.

Finally, the metadata ceiling could be lifted by adding features derived from app code, such as permissions exercised at runtime or static indicators of behavior, and a source code analysis. \citet{wang2018} categorize app removals according to reasons that metadata cannot observe. Features that describe what an app does, rather than what it declares, would
give a detector a basis for separating removals that follow a policy violation from removals with unrelated causes such as voluntary withdrawal. The configuration of the dataset centered on the developer, which this work did not analyze, is also available for the same comparison.

\bibliographystyle{apacite}
\bibliography{literature_bachelorthesis}

\clearpage

\begin{appendices}
    \clearpage
\onecolumn
\appendix
\section{Supplementary Tables}

This appendix contains supplementary data tables and additional results that support the methodology and findings discussed in Section 4. Specifically, it includes the per-label external validation rates, the full-data removal results, the balanced-subset ensemble removal results, and the extended feature direction analysis.

\begin{sidewaystable}[p]
\centering
\caption{Flag rates per label inside the original dataset and inside each validated set's matching side (removed against the malicious side, stable against the benign side), with ratios to the original rate. \texttt{adj} is the ratio after age standardization on deciles of days since last update. Values below 1 indicate under-representation among confirmed-correct labels.}
\label{tab:flag-rates}
\resizebox{\textheight}{!}{%
\begin{tabular}{llrr rrr rrr rrr rrr rrr}
\toprule
& & & & \multicolumn{3}{c}{T1} & \multicolumn{3}{c}{T2} & \multicolumn{3}{c}{T3} & \multicolumn{3}{c}{T4} & \multicolumn{3}{c}{T5} \\
\cmidrule(lr){5-7} \cmidrule(lr){8-10} \cmidrule(lr){11-13} \cmidrule(lr){14-16} \cmidrule(lr){17-19}
Set & Side & $n$ flagged & pct\_orig & pct & ratio & adj & pct & ratio & adj & pct & ratio & adj & pct & ratio & adj & pct & ratio & adj \\
\midrule
IF default ($c=0.05$) & removed & 24{,}439 & 5.00 & 5.76 & 1.15 & 1.17 & 4.88 & 0.98 & 1.01 & 4.53 & 0.91 & 0.94 & 4.53 & 0.91 & 0.93 & 4.63 & 0.92 & 0.94 \\
IF harsh ($c=0.20$) & removed & 97{,}753 & 20.00 & 21.50 & 1.07 & 1.09 & 19.09 & 0.95 & 0.99 & 17.81 & 0.89 & 0.92 & 17.86 & 0.89 & 0.91 & 18.31 & 0.92 & 0.93 \\
PI & removed & 117{,}117 & 23.96 & 20.38 & 0.85 & 0.83 & 16.71 & 0.70 & 0.65 & 15.11 & 0.63 & 0.59 & 14.44 & 0.60 & 0.58 & 13.88 & 0.58 & 0.57 \\
ND default ($k=5$) & removed & 107{,}788 & 22.05 & 17.84 & 0.81 & 0.80 & 15.63 & 0.71 & 0.68 & 14.75 & 0.67 & 0.65 & 14.33 & 0.65 & 0.63 & 14.08 & 0.64 & 0.63 \\
overlap default & removed & 3{,}309 & 0.68 & 0.55 & 0.81 & 0.78 & 0.44 & 0.64 & 0.60 & 0.38 & 0.57 & 0.53 & 0.32 & 0.47 & 0.45 & 0.29 & 0.43 & 0.42 \\
overlap harsh $k=5$ & removed & 10{,}886 & 2.23 & 1.65 & 0.74 & 0.72 & 1.27 & 0.57 & 0.53 & 1.12 & 0.50 & 0.47 & 1.00 & 0.45 & 0.43 & 0.95 & 0.42 & 0.42 \\
overlap harsh $k=50$ & removed & 13{,}662 & 2.80 & 2.26 & 0.81 & 0.78 & 1.82 & 0.65 & 0.61 & 1.63 & 0.58 & 0.54 & 1.50 & 0.54 & 0.52 & 1.45 & 0.52 & 0.51 \\
union default & removed & 189{,}585 & 38.79 & 34.51 & 0.89 & 0.88 & 29.84 & 0.77 & 0.74 & 27.78 & 0.72 & 0.69 & 27.12 & 0.70 & 0.68 & 26.57 & 0.69 & 0.68 \\
union harsh $k=5$ & removed & 238{,}503 & 48.80 & 45.81 & 0.94 & 0.93 & 40.47 & 0.83 & 0.81 & 37.86 & 0.78 & 0.76 & 37.37 & 0.77 & 0.76 & 37.27 & 0.76 & 0.76 \\
union harsh $k=50$ & removed & 225{,}300 & 46.10 & 43.62 & 0.95 & 0.94 & 38.04 & 0.83 & 0.81 & 35.59 & 0.77 & 0.76 & 35.12 & 0.76 & 0.75 & 35.28 & 0.77 & 0.76 \\
\midrule
IF default ($c=0.05$) & stable & 19{,}088 & 5.00 & 7.96 & 1.59 & 1.05 & 7.98 & 1.60 & 1.06 & 7.96 & 1.59 & 1.05 & 7.96 & 1.59 & 1.05 & 7.96 & 1.59 & 1.05 \\
IF harsh ($c=0.20$) & stable & 76{,}350 & 20.00 & 30.50 & 1.53 & 1.10 & 30.31 & 1.52 & 1.10 & 30.30 & 1.52 & 1.10 & 30.30 & 1.51 & 1.10 & 30.30 & 1.52 & 1.10 \\
PI & stable & 147{,}577 & 38.66 & 61.60 & 1.59 & 1.00 & 61.09 & 1.58 & 1.00 & 61.07 & 1.58 & 1.00 & 61.08 & 1.58 & 1.00 & 61.06 & 1.58 & 1.00 \\
ND default ($k=5$) & stable & 124{,}337 & 32.57 & 44.25 & 1.36 & 0.99 & 43.98 & 1.35 & 0.99 & 43.96 & 1.35 & 0.99 & 43.99 & 1.35 & 0.99 & 44.01 & 1.35 & 0.99 \\
overlap default & stable & 4{,}289 & 1.12 & 3.95 & 3.51 & 1.08 & 3.93 & 3.50 & 1.08 & 3.93 & 3.50 & 1.08 & 3.92 & 3.49 & 1.08 & 3.92 & 3.49 & 1.08 \\
overlap harsh $k=5$ & stable & 17{,}089 & 4.48 & 12.63 & 2.82 & 1.04 & 12.50 & 2.79 & 1.03 & 12.49 & 2.79 & 1.03 & 12.49 & 2.79 & 1.03 & 12.49 & 2.79 & 1.04 \\
overlap harsh $k=50$ & stable & 20{,}108 & 5.27 & 14.99 & 2.85 & 1.06 & 14.86 & 2.82 & 1.05 & 14.86 & 2.82 & 1.06 & 14.86 & 2.82 & 1.06 & 14.85 & 2.82 & 1.06 \\
union default & stable & 203{,}369 & 53.27 & 71.84 & 1.35 & 1.00 & 71.40 & 1.34 & 1.00 & 71.39 & 1.34 & 1.00 & 71.40 & 1.34 & 1.00 & 71.40 & 1.34 & 1.00 \\
union harsh $k=5$ & stable & 230{,}308 & 60.33 & 77.76 & 1.29 & 1.01 & 77.32 & 1.28 & 1.01 & 77.32 & 1.28 & 1.01 & 77.33 & 1.28 & 1.01 & 77.33 & 1.28 & 1.01 \\
union harsh $k=50$ & stable & 224{,}592 & 58.83 & 76.71 & 1.30 & 1.02 & 76.27 & 1.30 & 1.01 & 76.27 & 1.30 & 1.01 & 76.28 & 1.30 & 1.01 & 76.27 & 1.30 & 1.01 \\
\bottomrule
\end{tabular}%
}
\end{sidewaystable}

\begin{table*}[htbp]
\centering
\footnotesize
\caption{Full-data removal results. Each row removes the flagged apps from the training set of 609,359 and retrains the model. $\Delta$AUC is measured against the baseline of 0.7927.}
\label{tab:removal_full}
\begin{tabular}{lrrrrrr}
\toprule
Model & Removed & AUC & $\Delta$AUC & P & R & F1 \\
\midrule
Baseline (no removal) & 0 & 0.7927 & --- & 0.7661 & 0.7053 & 0.7345 \\
\addlinespace
IF $c=0.01$ & 6,069 & 0.7909 & $-0.0018$ & 0.7654 & 0.7036 & 0.7332 \\
IF $c=0.05$ (default) & 30,430 & 0.7909 & $-0.0018$ & 0.7660 & 0.7021 & 0.7327 \\
IF $c=0.10$ & 60,757 & 0.7891 & $-0.0036$ & 0.7636 & 0.7027 & 0.7319 \\
IF $c=0.15$ & 91,189 & 0.7869 & $-0.0058$ & 0.7601 & 0.7052 & 0.7316 \\
IF $c=0.20$ (harsh) & 121,774 & 0.7825 & $-0.0102$ & 0.7557 & 0.7070 & 0.7305 \\
\addlinespace
ND $k=5$ (default) & 162,690 & 0.7879 & $-0.0048$ & 0.7577 & 0.7191 & 0.7379 \\
ND $k=10$ & 164,084 & 0.7867 & $-0.0060$ & 0.7635 & 0.7032 & 0.7321 \\
ND $k=20$ & 164,071 & 0.7835 & $-0.0092$ & 0.7557 & 0.7145 & 0.7345 \\
ND $k=50$ (harsh) & 169,062 & 0.7796 & $-0.0131$ & 0.7478 & 0.7297 & 0.7386 \\
\addlinespace
PI & 185,451 & 0.7695 & $-0.0232$ & 0.7267 & 0.7491 & 0.7378 \\
\addlinespace
Overlap default & 5,288 & 0.7924 & $-0.0003$ & 0.7662 & 0.7024 & 0.7329 \\
Overlap harsh $k=5$ & 19,577 & 0.7905 & $-0.0022$ & 0.7642 & 0.7045 & 0.7331 \\
Overlap harsh $k=50$ & 23,603 & 0.7891 & $-0.0036$ & 0.7636 & 0.7021 & 0.7316 \\
\addlinespace
Union default & 275,254 & 0.7709 & $-0.0218$ & 0.7278 & 0.7517 & 0.7396 \\
Union harsh $k=5$ & 328,291 & 0.7699 & $-0.0228$ & 0.7258 & 0.7533 & 0.7393 \\
Union harsh $k=50$ & 315,210 & 0.7699 & $-0.0228$ & 0.7262 & 0.7534 & 0.7395 \\
\bottomrule
\end{tabular}
\end{table*}

\begin{table*}[htbp]
\centering
\footnotesize
\caption{Label flipping. The labels of the overlap set are flipped in the training data and the model retrained. The top block reports the number of apps flipped and the result at the 100,000 training size against a baseline of 0.7941. The bottom block reports test AUC as the training size grows. No setting moves the model beyond the range of the baseline.}
\label{tab:flip}
\begin{tabular}{lrrrrrr}
\toprule
Model & Flipped & AUC-val & AUC-test & P & R & F1 \\
\midrule
Baseline (no flip) & 0 & 0.7927 & 0.7941 & 0.7666 & 0.7064 & 0.7352 \\
Flip overlap default & 3,021 & 0.7924 & 0.7935 & 0.7665 & 0.7053 & 0.7346 \\
Flip overlap harsh ($k=5$) & 11,120 & 0.7931 & 0.7942 & 0.7656 & 0.7098 & 0.7367 \\
Flip overlap harsh ($k=50$) & 13,395 & 0.7877 & 0.7886 & 0.7621 & 0.7065 & 0.7333 \\
\bottomrule
\end{tabular}

\vspace{1em}

\begin{tabular}{lrrrrrr}
\toprule
Model & 2K & 5K & 10K & 25K & 50K & 100K \\
\midrule
Baseline & 0.7742 & 0.7851 & 0.7880 & 0.7924 & 0.7938 & 0.7941 \\
Flip overlap default & 0.7749 & 0.7818 & 0.7862 & 0.7912 & 0.7885 & 0.7935 \\
Flip overlap harsh ($k=5$) & 0.7762 & 0.7820 & 0.7849 & 0.7888 & 0.7901 & 0.7942 \\
Flip overlap harsh ($k=50$) & 0.7740 & 0.7805 & 0.7843 & 0.7887 & 0.7881 & 0.7886 \\
\bottomrule
\end{tabular}
\end{table*}

\begin{table*}[htbp]
\centering
\footnotesize
\caption{Ensemble removal cross-check at the 100,000 training size. Removal is repeated using the balanced-subset ensemble of the baseline, with the validation set drawn once and held fixed. $\Delta$AUC is measured against the ensemble baseline of 0.7941. The pattern matches the full-data removal, so the effect does not depend on the training procedure.}
\label{tab:removal_ensemble}
\begin{tabular}{lrrrrr}
\toprule
Model & AUC & $\Delta$AUC & P & R & F1 \\
\midrule
Baseline (no removal) & 0.7941 & --- & 0.7666 & 0.7064 & 0.7352 \\
\addlinespace
Removed IF default ($c=0.05$) & 0.7923 & $-0.0018$ & 0.7664 & 0.7036 & 0.7336 \\
Removed IF harsh ($c=0.20$) & 0.7876 & $-0.0065$ & 0.7588 & 0.7120 & 0.7347 \\
Removed ND default ($k=5$) & 0.7906 & $-0.0035$ & 0.7593 & 0.7234 & 0.7409 \\
Removed PI & 0.7689 & $-0.0252$ & 0.7263 & 0.7511 & 0.7385 \\
Removed union default & 0.7703 & $-0.0238$ & 0.7266 & 0.7540 & 0.7400 \\
Removed union harsh ($k=5$) & 0.7691 & $-0.0250$ & 0.7247 & 0.7549 & 0.7395 \\
Removed union harsh ($k=50$) & 0.7693 & $-0.0248$ & 0.7250 & 0.7566 & 0.7404 \\
\bottomrule
\end{tabular}
\end{table*}

\begin{table*}[htbp]
\centering
\footnotesize
\caption{Class-averaged values for the top features from the baseline model, computed on the training set. The final column reports which label carries the higher mean value.}
\label{tab:feature_directions}
\begin{tabular}{lrrl}
\toprule
Feature & Mean (removed) & Mean (stable) & Higher in \\
\midrule
days\_since\_last\_update & 614.35 & 370.70 & removed \\
ContentRating\_Unrated & 0.2050 & 0.0295 & removed \\
privacy\_policy\_link & 0.7609 & 0.5500 & removed \\
isSpamming & 0.7524 & 0.5882 & removed \\
developer\_address & 0.6909 & 0.5897 & removed \\
developer\_website & 0.4309 & 0.3342 & removed \\
DeveloperCategory\_Moderate & 0.4493 & 0.3719 & removed \\
SMS & 0.0420 & 0.0168 & removed \\
Genre\_Casual & 0.0448 & 0.0282 & removed \\
Genre\_Personalization & 0.0621 & 0.0484 & removed \\
Genre\_Entertainment & 0.0852 & 0.0636 & removed \\
AndroidVersion\_2.3 and up & 0.2163 & 0.1947 & removed \\
PHONE & 0.2737 & 0.2565 & removed \\
max\_downloads\_log & 6.8726 & 8.0168 & stable \\
ContentRating\_Everyone & 0.7056 & 0.8800 & stable \\
LenWhatsNew & 1.5916 & 2.1313 & stable \\
DeveloperCategory\_Aggressive & 0.0874 & 0.1292 & stable \\
CONTACTS & 0.1745 & 0.1786 & stable \\
media & 0.0115 & 0.0202 & stable \\
paid & 0.0005 & 0.0027 & stable \\
\bottomrule
\end{tabular}
\end{table*}

\begin{figure}[htbp]
\centering
\includegraphics[width=\columnwidth]{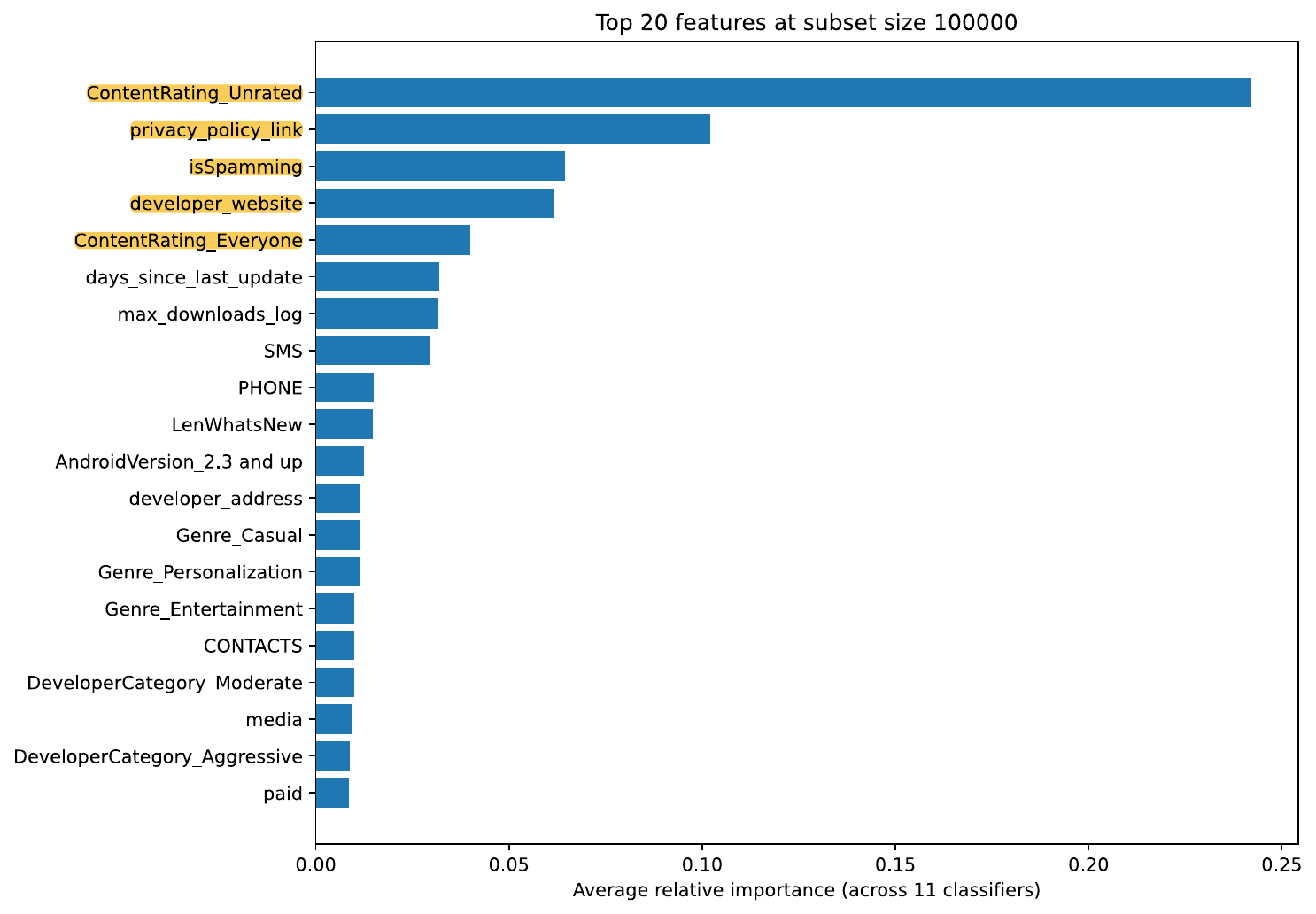}
\caption{The twenty most important features of the baseline model, averaged across the 11 classifiers of the ensemble at a subset size of 100,000. Importance is reported as an average relative value. The presence of a link to a privacy policy ranks second and the presence of a field for a developer website ranks fourth.}
\label{fig:importance}
\end{figure}

\clearpage
\section{Reproducibility}
\label{app:repro}

This appendix records the data, the parameters, the random seeds and the scripts behind
every result, so that the pipeline can be rerun or extended.

\subsection*{Data}

All experiments use the corrected dataset and the five validated datasets published by
\citet{mohsen2026data} at \url{https://doi.org/10.34894/XV84Z8}. Both are reproduced in the
accompanying material.

\begin{description}
  \item[\texttt{original\_dataset\_corrected.csv}.] The full population of 870,514 apps in
    the configuration centered on the user, which holds the 47 features listed in
    Appendix~C together with the binary \texttt{status} label and the \texttt{pkgname}
    identifier. Of these apps, 56.1\% carry the removed label.
  \item[\texttt{val\_dataset\_t1.csv} through \texttt{val\_dataset\_t5.csv}.] The five
    validated subsets described in Section~3.8, each containing the apps whose status
    agrees with both the VirusTotal and the Quark Engine verdict. Every set holds a
    malicious side and a benign side rather than malicious apps alone. Confidence in the
    removed label increases with the index, from $T_1$ with 62,093 removed and 18,275
    stable apps to $T_5$ with 19,027 removed and 19,275 stable apps, while the benign
    criterion loosens over the same range. The sets are matched against the flagged apps
    by \texttt{pkgname}.
\end{description}

\subsection*{Environment}

Experiments were run on the H\'abr\'ok high performance cluster with Python~3.11.3,
\texttt{scikit-learn}, \texttt{xgboost}, \texttt{pandas} and \texttt{numpy}, on 16~cores with
64~GB of memory. Every random seed is fixed at $42$, so each script reproduces its results
exactly in the same environment. Exact package versions are recorded in the job environment
on H\'abr\'ok and are available on request.

\subsection*{Preprocessing and partitions}

The \texttt{Unnamed:~0} and \texttt{pkgname} columns are dropped. \texttt{CurrentVersion} is
converted with \texttt{pandas.to\_numeric(errors="coerce")}, which turns non-numeric version
strings into missing values. The remaining categorical features are expanded with
\texttt{get\_dummies(drop\_first=True)}, giving 333 columns. The outer split is
\texttt{train\_test\_split(test\_size=0.30, stratify=status, random\_state=42)}, giving
609,359~training and 261,155~test apps. The validation set is drawn once from the
training set with \texttt{test\_size=0.4285, random\_state=42}, giving 261,111~validation
apps and leaving a trainable partition of 348,248. The test set is never modified by
any intervention. Isolation Forest and Neighborhood Disagreement operate on features that are
median-imputed with \texttt{SimpleImputer} and standardized with \texttt{StandardScaler}, the
XGBoost models operate on the unscaled encoded features.

\subsection*{Detector parameters}

\begin{description}
  \item[Isolation Forest.] \texttt{n\_estimators=100}, \texttt{max\_samples=50000},
    \texttt{random\_state=42}, contamination $c \in \{0.01, 0.05, 0.10, 0.15, 0.20\}$. The
    detector is fitted separately within each label class, so that anomaly means unusual
    relative to other apps carrying the same label.
  \item[Neighborhood Disagreement.] Euclidean nearest neighbors on the scaled features, with
    the index built once at $k = 50$ and sliced for $k \in \{5, 10, 20, 50\}$. An app
    is flagged when the majority label of its $k$ nearest neighbors, excluding itself,
    differs from its own, ties are resolved toward the stable label.
  \item[Prediction Inconsistency.] \texttt{LogisticRegression(max\_iter=1000)} with
    \texttt{StratifiedKFold(n\_splits=5, shuffle=True, random\_state=42)} and
    \texttt{cross\_val\_predict}, so that every app is scored by a model trained on
    the other four folds. Logistic regression is deliberately a different model family from
    the XGBoost predictor, to avoid using the predictor to clean its own training data.
\end{description}

\subsection*{Model parameters}

The replication of \citet{mohsen2022} trains 11 \texttt{XGBClassifier} models on balanced
subsets, with \texttt{n\_estimators} drawn from $\{256, 512\}$ and \texttt{max\_depth} from
$\{2, 3\}$ using \texttt{random.seed(42)}, \texttt{tree\_method="hist"} and
\texttt{eval\_metric="logloss"}. The Validation~1 experiments use the same ensemble but train
each classifier on the entire training set, with
\texttt{scale\_pos\_weight}~$=n_{\text{stable}}/n_{\text{removed}}$ replacing the balanced
subsets. The single full-data model used for the parameter sweeps in Table~A.2 uses
\texttt{n\_estimators=512}, \texttt{max\_depth=3}, \texttt{learning\_rate=0.1} and the same
\texttt{scale\_pos\_weight}. The hyperparameter sequence is regenerated from the same seed in
every scenario, so rows within a table are directly comparable.

\subsection*{Supporting files}

Three files in the accompanying material produce no result of their own but are required in
order to run or to interpret the pipeline.

\begin{description}
  \item[\texttt{thesis\_pipeline.py}.] A shared module that holds the
    configuration, the dataset and its four partitions, the three label noise detectors, the
    samplers, the model trainers and the evaluation metrics. Every experiment script imports
    from it, so the encoding, the splits and the training procedure are defined once and the
    scripts differ only in the intervention they apply. It is never run directly.
  \item[\texttt{mohsen\_baseline\_original.py}.] The original implementation of the removal
    prediction model of \citet{mohsen2022}, included unmodified. It is the code the baseline
    of Section~4.1 reproduces, so it allows the replication to be checked against the
    implementation itself rather than against the published description alone.
  \item[\texttt{README.md}.] Records the environment, the order in which the scripts are to
    be run, the expected flag counts and the provenance of the stored flagged sets.
\end{description}

\subsection*{Scripts and outputs}

Each flagged set is stored as a list of row indices into the original dataset, so any
experiment can be rerun without recomputing the detectors. The \texttt{\_v3} suffix marks the
current version of each flagged-set file, earlier versions were superseded by the leakage
correction described in Section~4.4 and are not used. The detector scripts are therefore run
first, because every later script reads the stored flagged sets rather than recomputing them.
The only exception is \texttt{step4\_directions.py}, which reads the dataset alone.
Table~\ref{tab:repro} maps each script to its output and to the place in this work where the output appears, in the order in which the scripts are run.

\begin{table}[!ht]
\centering
\caption{Scripts, their outputs, and where each result appears.}
\label{tab:repro}
\small
\begin{tabular}{@{}p{0.26\linewidth}p{0.40\linewidth}p{0.28\linewidth}@{}}
\hline
Script & Output & Appears in \\
\hline
\texttt{step4bagain.py}       & \texttt{flagged\_IF/ND/PI\_*\_v3.csv}      & Section 4.2, Figure 2 \\
\texttt{step4bpi.py}          & \texttt{flagged\_overlap/union\_*\_v3.csv} & Section 4.2, Figure 2 \\
\texttt{v1\_pipeline.py}      & \texttt{v1\_results.csv}                   & Section 4.3, Table 1 \\
\texttt{step5\_fulltrain.py}  & \texttt{full\_removal\_v3.csv}             & Table 5, Figure 3 \\
\texttt{step4d\_removal.py}   & \texttt{removal\_sweep\_v3.csv}            & Table 7 \\
\texttt{step4e\_flip.py}      & \texttt{flip\_sweep\_v3.csv}               & Table 6 \\
\texttt{step4f\_trainedon.py} & \texttt{trainedon\_v3.csv}                 & Table 2, Figure 4 \\
\texttt{step4g\_mirror.py}    & \texttt{mirror\_v3.csv}                    & Table 3 \\
\texttt{step4\_directions.py} & \texttt{feature\_directions.csv}           & Table 8 \\
\texttt{v2\_final\_table.py}  & \texttt{v2\_final\_table.csv}              & Table 4 \\
\hline
\end{tabular}
\end{table}

\FloatBarrier
\subsection*{Continuing this work}

Three extensions can be run directly on the stored flagged sets, without recomputing any
detector. Targeted relabeling can use the two directions reported by the mirror analysis
instead of flipping every app in the overlap. Confident Learning
\citep{northcutt2021cl} can replace the binary flag with a per-app probability of
mislabeling. And the complement of the validated datasets, the apps whose status the
two external tools contradict, would give direct external mislabel candidates to intersect
with the flagged sets.
 
\clearpage
\section{Feature List}
\label{app:features}

Table~\ref{tab:features} lists the 47 features of the user-centered configuration of the
dataset, as they appear in \texttt{original\_dataset\_corrected.csv}. They fall into six
groups: how an app presents itself in the store, the ratings it has received, how
recently it has been updated and which Android versions it supports, what its developer
discloses, which permission groups it requests, and which Android API namespaces its
manifest refers to. The label \texttt{status} and the identifier \texttt{pkgname} are not
features and are excluded from this count.

Five of the 47 are categorical: \texttt{Genre}, \texttt{ContentRating},
\texttt{AndroidVersion}, \texttt{DeveloperCategory} and \texttt{lowest\_android\_version}.
These are expanded into binary indicator columns with
\texttt{get\_dummies(drop\_first=True)} before modeling, which is what raises the feature
space from 47 columns to 333.

\begin{table}[!ht]
\centering
\caption{The 47 features of the user-centered configuration, grouped by what they describe.}
\label{tab:features}
\small
\begin{tabular}{@{}p{0.24\linewidth}p{0.72\linewidth}@{}}\hline
Group & Features \\
\hline
Store listing &
\texttt{LenTitle}, \texttt{LenDescription}, \texttt{LenWhatsNew}, \texttt{Genre},
\texttt{ContentRating}, \texttt{paid}, \texttt{file\_size},
\texttt{max\_downloads\_log} \\[2pt]
Ratings and reviews &
\texttt{ReviewsAverage}, \texttt{OneStarRatings}, \texttt{TwoStarRatings},
\texttt{ThreeStarRatings}, \texttt{FourStarRatings}, \texttt{FiveStarRatings} \\[2pt]
Versioning and updates &
\texttt{CurrentVersion}, \texttt{LastUpdated}, \texttt{days\_since\_last\_update},
\texttt{AndroidVersion}, \texttt{lowest\_android\_version},
\texttt{highest\_android\_version} \\[2pt]
Developer &
\texttt{DeveloperCategory}, \texttt{DevRegisteredDomain}, \texttt{developer\_email},
\texttt{developer\_website}, \texttt{developer\_address}, \texttt{privacy\_policy\_link},
\texttt{isSpamming} \\[2pt]
Permission groups &
\texttt{CALENDAR}, \texttt{CAMERA}, \texttt{CONTACTS}, \texttt{LOCATION},
\texttt{MICROPHONE}, \texttt{PHONE}, \texttt{SENSORS}, \texttt{SMS},
\texttt{STORAGE} \\[2pt]
API namespaces &
\texttt{net}, \texttt{intent}, \texttt{bluetooth}, \texttt{app}, \texttt{provider},
\texttt{speech}, \texttt{nfc}, \texttt{media}, \texttt{hardware}, \texttt{google},
\texttt{os} \\
\hline
\end{tabular}
\end{table}
    \newpage
\end{appendices}

\end{document}